\documentclass[a4paper,11pt]{article}
\usepackage{jheppub} 
\usepackage[T1]{fontenc} 
\usepackage[capitalize]{cleveref}

\title{IceCube and the origin of ANITA-IV events}

\author[a]{Toni Bert\'olez-Mart\'inez,}
\author[b]{Carlos A. Arg\"uelles,}
\author[c,d]{Ivan Esteban,}
\author[e]{Jacobo Lopez-Pavon,}
\author[b]{Ivan Martinez-Soler,}
\author[a]{Jordi Salvado}

\affiliation[a]{Departament de F\'isica Qu\`antica i Astrof\'isica and Institut de Ci\`encies del Cosmos, Universitat de Barcelona, Diagonal 647, E-08028 Barcelona, Spain}
\affiliation[b]{Department of Physics \& Laboratory for Particle Physics and Cosmology, Harvard University, Cambridge, MA 02138, USA}
\affiliation[c]{Center for Cosmology and AstroParticle Physics (CCAPP), Ohio State University, Columbus, OH 43210}
\affiliation[d]{Department of Physics, Ohio State University, Columbus, OH 43210}
\affiliation[e]{Instituto de F\'isica Corpuscular, Universidad de Valencia and CSIC, Edificio Insitutos Investigaci\'on, Catedr\'atico Jos\'e Beltr\'an 2, 46980, Paterna, Spain}

\emailAdd{antoni.bertolez@fqa.ub.edu}
\emailAdd{esteban.6@osu.edu}
\emailAdd{carguelles@g.harvard.edu}
\emailAdd{jlpavon@ific.uv.es}
\emailAdd{imartinezsoler@fas.harvard.edu}
\emailAdd{jor.salvado@gmail.com}

\abstract{%
Recently, the ANITA collaboration announced the detection of new, unsettling upgoing Ultra-High-Energy (UHE) events. 
Understanding their origin is pressing to ensure success of the incoming UHE neutrino program.
In this work, we study their internal consistency and the implications of the lack of similar events in IceCube.
We introduce a generic, simple parametrization to study the compatibility between these two observatories in Standard Model-like and Beyond Standard Model scenarios: an incoming flux of particles that interact with Earth nucleons with cross section $\sigma$, producing particle showers along with long-lived particles that decay with lifetime $\tau$ and generate a shower that explains ANITA observations. 
We find that the ANITA angular distribution imposes significant constraints, and when including null observations from IceCube only $\tau \sim 10^{-3}$--$10^{-2} \,\mathrm{s}$ and $\sigma \sim 10^{-33}$ --$10^{-32}\,\mathrm{cm^2}$ can explain the data.
This hypothesis is testable with future IceCube data. Finally, we discuss a specific model that can realize this scenario.
Our analysis highlights the importance of simultaneous observations by high-energy optical neutrino telescopes and new UHE radio detectors to uncover cosmogenic neutrinos or discover new physics.
}

\newcommand{\Nu}{\textrm{N}}
\newcommand{\Tau}{\textrm{T}}

\begin{document} 

\preprint{\hfill FTUV-23-0426.7835; IFIC/23-14}
\maketitle
\flushbottom

\section{Introduction}

It has been over a decade since the IceCube Neutrino Observatory, located at the geographical South Pole, opened a new window to observe the Universe by detecting high-energy astrophysical neutrinos with energies up to 6\,PeV~\cite{IceCube:2021rpz}.
This decade has witnessed the progress from observing a flux compatible with an isotropic distribution~\cite{IceCube:2013low}, the so-called diffuse flux; to the discovery of first sources~\cite{IceCube:2018dnn, IceCube:2022der}.
The confidence of these observations as astrophysical neutrinos is extremely high, not only because the sample sizes have grown to hundreds over the last decade, but also because systematic uncertainties are controlled by in-situ measurements.
IceCube first measured the atmospheric neutrino flux~\cite{IceCube:2010whx}, proving that it could constrain its backgrounds and that the event reconstructions were reliable, and then discovered the astrophysical component on top of this one.

As we venture into this new decade, radio detectors place themselves as a promising technology to extend the observations of IceCube to Ultra-High Energies (UHE). 
This requires surmounting significant technological and logistical challenges.
Among the challenges, one stands out.
Unlike IceCube or ANTARES~\cite{ANTARES:2011hfw}, where the detectors can \textit{calibrate} and \textit{test} their selection procedures and reconstructions on the well-understood atmospheric neutrino flux, in this upcoming generation such calibration is much more challenging.
Fortunately, the sensitivity of present UHE experiments is comparable with IceCube limits, so current observations can be cross-checked with existing or future IceCube data.

\begin{figure}[b]
	\centering
	\includegraphics[width=0.495\linewidth]{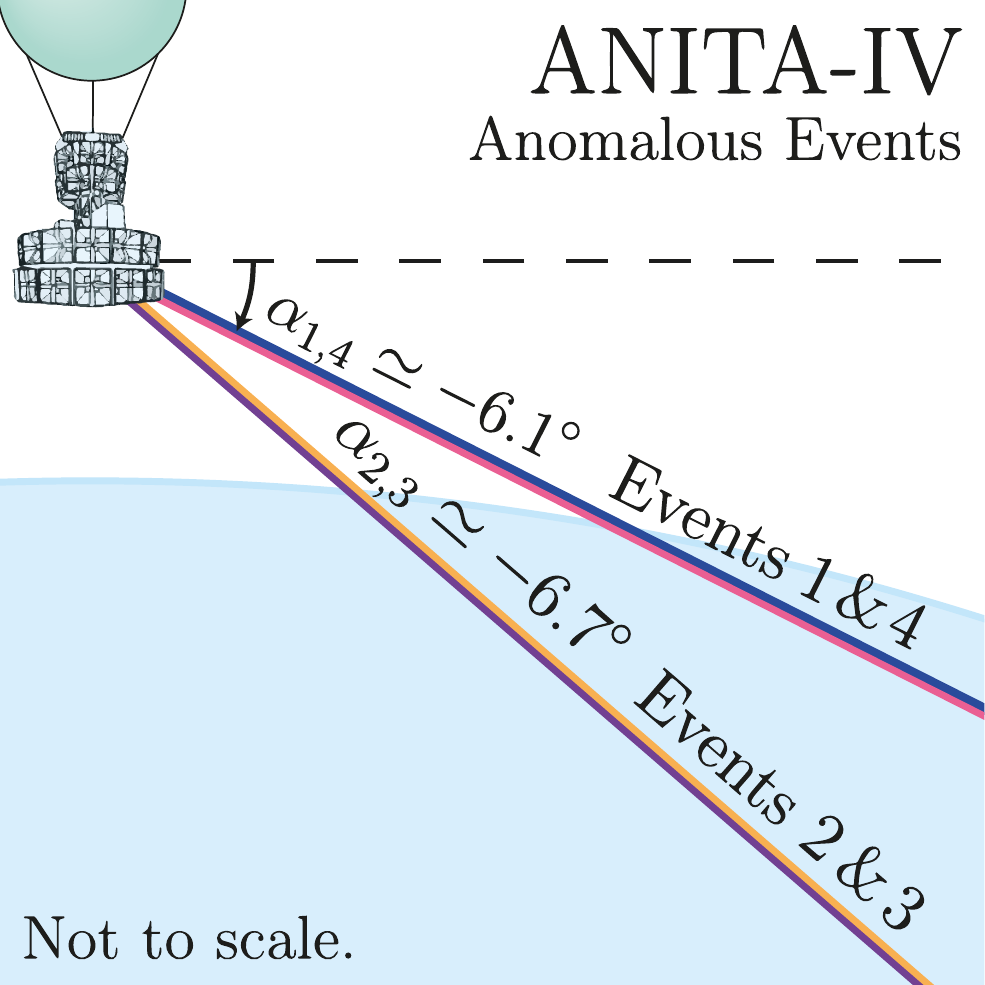}
 	\includegraphics[width=0.495\linewidth]{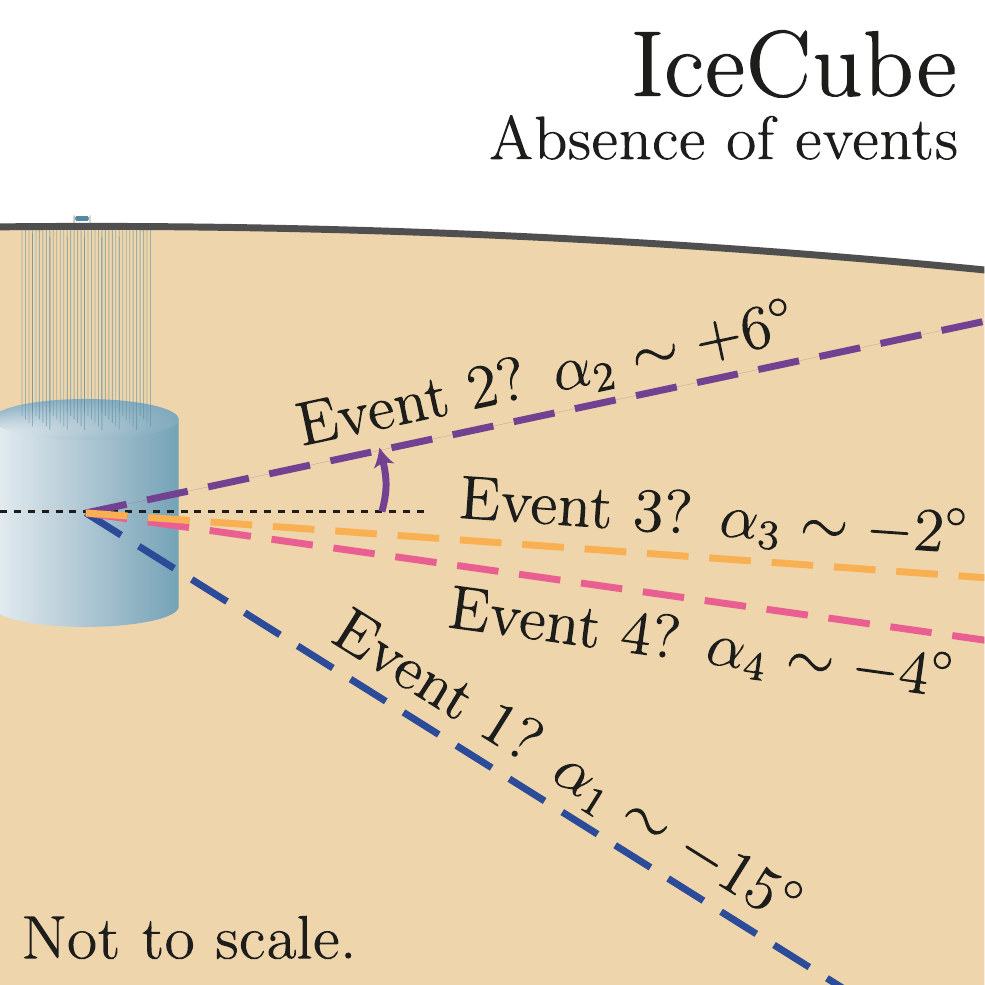}
	\caption{Directions of the Anomalous Events  as viewed from ANITA (left) and IceCube (right). \emph{Particle physics models that explain the ANITA observations are challenged by the absence of signals at IceCube}. }
	\label{fig:aaes}
\end{figure}

First results are showing up. The ANITA-IV experiment, a balloon flying over Antarctica, has observed four events with energies $\sim$ 1\,EeV and incident directions $\sim$ $1^\circ$ below the horizon with a significance $\sim 3\sigma$~\cite{ANITA:2020gmv, ANITA:2021xxh}, that we depict in \Cref{fig:aaes}.
The directions are compatible with a neutrino origin, which would make them the highest-energy neutrinos ever observed.
However, a Standard Model (SM) explanation is in tension with non-observations from Auger~\cite{ANITA:2021xxh} and, as we show below, IceCube.

It is pressing to understand the origin of these events, whether novel physics or background, in order to guarantee the success of other UHE detectors with better sensitivity than ANITA.
If they are background, further work is needed to understand their origin and filter them out to avoid overwhelming the neutrino signal.
If they have a Beyond the Standard Model (BSM) origin, we must robustly understand which particle models can explain them and their predictions in different detectors, to fully confirm this hypothesis.

In this article, we exploit the aforementioned connection with IceCube to diagnose if these detections correspond to neutrinos or to some BSM scenario, focusing on the latter. By exploring an energy scale never reached before with fundamental particles, ANITA is opening a new window to such scenarios~\cite{Denton:2020jft, Huang:2021mki, Valera:2022ylt, Esteban:2022uuw}. We perform a model-independent analysis, leaving most details to model builders, but we briefly comment on specific models and encourage the reader to read Refs.~\cite{Fox:2018syq, Chauhan:2018lnq, Heurtier:2019git, Hooper:2019ytr, Anchordoqui:2019utb, Anchordoqui:2018ucj, Dudas:2018npp, Huang:2018als, Connolly:2018ewv, Cherry:2018rxj, Anchordoqui:2018ssd, Collins:2018jpg, Esteban:2019hcm, Chipman:2019vjm, Abdullah:2019ofw, Borah:2019ciw, Esmaili:2019pcy, 
Heurtier:2019rkz, Altmannshofer:2020axr, Cline:2019snp, Liang:2021rnv} for proposed models on UHE anomalous events.

Our method extends beyond ANITA and IceCube, since these are expected to be accompanied soon by a family of optical --- such as KM3NeT~\cite{KM3Net:2016zxf}, P-ONE~\cite{P-ONE:2020ljt}, or Baikal-GVD~\cite{Baikal-GVD:2018isr} ---, radio --- such as PUEO~\cite{PUEO:2020bnn}, GRAND~\cite{GRAND:2018iaj}, TAROGE~\cite{TAROGE:2022soh}, BEACON~\cite{Wissel:2020fav}, RET~\cite{RadarEchoTelescope:2021rca}, or RNO-G~\cite{RNO-G:2020rmc} ---, or even acoustic neutrino detectors --- such as ANDIAMO~\cite{Marinelli:2021upw}.
We do not discuss here the connection with Earth-skimming experiments --- such as TAMBO~\cite{Romero-Wolf:2020pzh}, TRINITY~\cite{Otte:2019knb} or POEMMA~\cite{POEMMA:2020ykm} ---, since they have not yet reached the sensitivity to detect neutrinos, but we expect them to also provide important information as they look at the region of Earth that is mostly transparent to neutrinos. 

The rest of this article is organized as follows.
In~\cref{sec:bsm-expla}, we introduce our model-independent parametrization of generic BSM models; in~\cref{sec:steady-or-not} we discuss the source properties; in~\cref{sec:anita-angular} we study the inner consistency of ANITA's data; in~\cref{sec:icecube-interplay} we report on the consistency between ANITA and IceCube observations; in~\cref{sec:models} we give the reader a brief discussion of potential models; and in~\cref{sec:conclu} we conclude.

\section{Beyond the Standard Model explanation\label{sec:bsm-expla}}

In this section, following a model independent approach, we parametrize a class of BSM models that could explain the anomalous events detected by ANITA.
We seek generic modeling, showing that few parameters capture the main physics.

\begin{figure}[hbtp]
	\centering
	\includegraphics[width=0.495\linewidth]{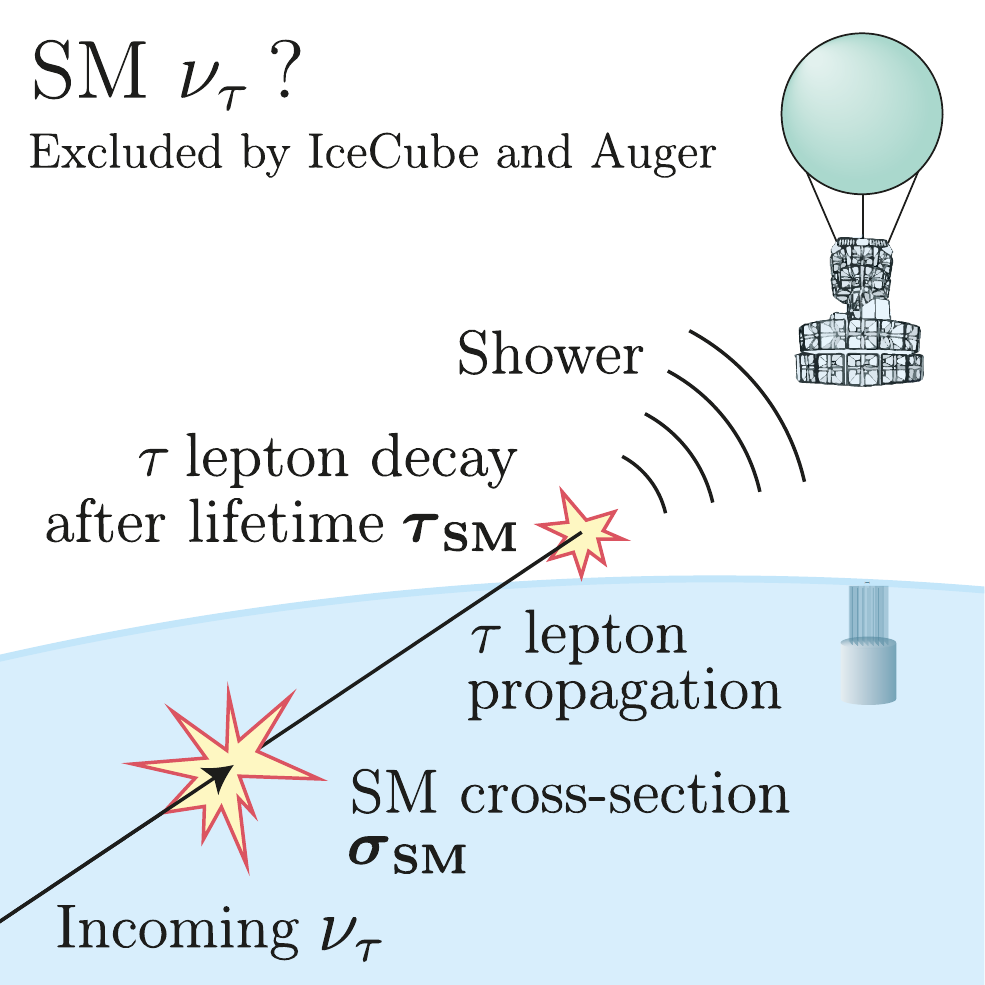}
 	\includegraphics[width=0.495\linewidth]{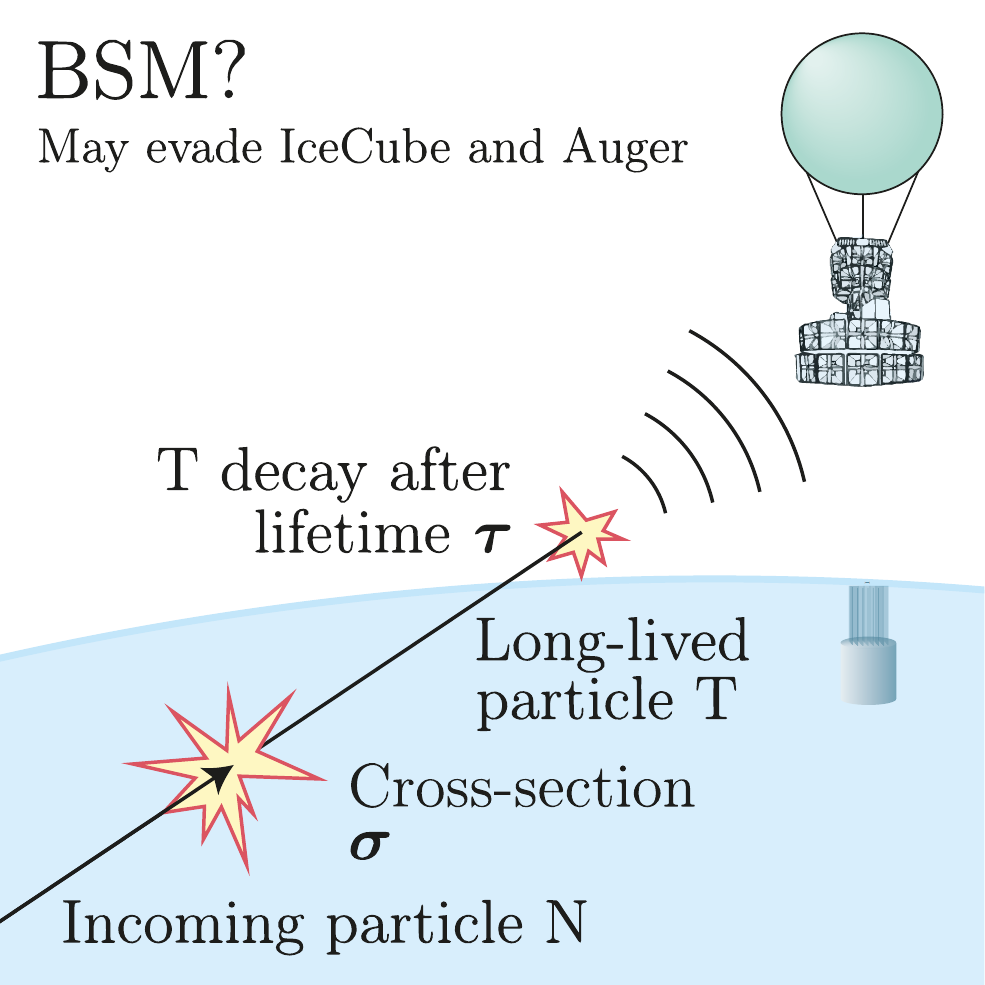}
	\caption{Possible origins of the ANITA anomalous events. In the SM, $\nu_\tau$ produce a $\tau$ lepton that decays in-air (left). We parametrize a class of BSM models by incoming particles that produce, with cross section $\boldsymbol{\sigma}$, long-lived particles that decay with lifetime $\boldsymbol{\tau}$ (right). \emph{Event distributions only depend on $\sigma$ and $\tau$}.}
	\label{fig:bsm}
\end{figure}

\Cref{fig:bsm} shows how anomalous events can be generated.
In the SM, an incoming $\nu_\tau$ flux interacts with Earth nucleons, producing $\tau$ leptons that decay and generate a shower observed by ANITA.
In BSM, we consider an incoming flux of particles, denoted as $\Nu$, that interact with Earth nucleons with cross section $\sigma$ producing long-lived particles, denoted as $\Tau$, that decay with lab-frame lifetime $\tau$ and generate a shower observed by ANITA. 

Our simplified BSM models are fully determined by three parameters: the incoming $\Nu$ flux, $\Phi$; the $\Nu$-nucleon interaction cross section, $\sigma$; and the lab-frame $\Tau$ lifetime, $\tau$.
When $\sigma$ is the SM neutrino-nucleon cross section and $\tau$ the $\tau$-lepton lifetime, this parametrization approximates the SM.
Although a SM explanation is inconsistent with Auger UHE neutrino limits~\cite{ANITA:2021xxh} (and IceCube, as we show below), different $\sigma$ and $\tau$ modify the event morphologies in those experiments, leading to potentially weaker constraints as we explore below.

This parametrization is a proof-of-principle of scenarios where particle showers produce the events.
It does not exhaustively explore all BSM models, and more detailed model-by-model studies can still be done; we comment on this in \cref{sec:models}. There are also models for previous UHE anomalous events that do not invoke particle showers~\cite{Esteban:2019hcm}, and hence do not fall under our parametrization.

We seek a minimal, conservative approach, assuming an incoming flux \emph{only} at the energies where ANITA has observed the anomalous events.
We also ignore effects that would redistribute particle energies --- including $\Tau$ or $\Nu$ energy losses, or different energies of $\Tau$ at production ---, as the detection efficiencies of IceCube and ANITA are quite flat at the energies we consider~\cite{ANITA:2021xxh,IceCube:2016uab}.
We assume T absorption by Earth with the production cross section $\sigma$, expected from the time-reversal invariance of the production process.
As we show in \cref{app:absorption}, our conclusions do not change if T is not absorbed by Earth. 

\subsection{Number of expected events}

There are four signals that the scenario we consider can produce. First, a shower is produced when $\Tau$ decays, which generates the signals at ANITA.
Second, a shower produced when $\Nu$ interacts with Earth to produce $\Tau$. (This signal can be avoided if the hadronic part of the interaction between $\Nu$ and nuclei to produce $\Tau$ is very elastic, which may happen for instance in models with light mediators~\cite{GarciaSoto:2022vlw}.)
Third, a shower produced if $\Tau$ gets absorbed by Earth.
Fourth, a track if $\Tau$ is charged.
Since the last signal can only be detected by IceCube and is easily avoided if $\Tau$ is electrically neutral, we conservatively ignore it.

The number of events per solid angle produced at elevation $\alpha$ by the decay of T is
\begin{equation}
    \frac{\mathrm{d}N_\Tau^\mathrm{dec}(\alpha)}{\mathrm{d}\Omega} =  \frac{\Phi}{4\pi} \, P^\Tau_\mathrm{exit}(\alpha) \, P^\Tau_\mathrm{decay}(\alpha) \, A_\mathrm{eff} \, \Delta t \, ,
    \label{eq:T_events}
\end{equation}
where $\Phi$ is the flux of the incoming $\Nu$ particles; $P^\Tau_\mathrm{exit}$ is the probability for N to hit a nucleon and produce a T that reaches the detector; $P^\Tau_\mathrm{decay}(\alpha) = 1 - e^{-d(\alpha)/\tau}$ is the probability for $T$ to decay inside the effective volume of size $d(\alpha)$ in the incoming direction $\alpha$; $A_\mathrm{eff}$ is the effective area to which the detector is sensitive, including detection efficiency; and $\Delta t$ is the whole observation period over which we integrate. 
We assume that the direction of the shower at ANITA corresponds to the incoming direction of $\Tau$ and $\Nu$, i.e., that all particles are relativistic. The full expression of $P^\Tau_\mathrm{exit}(\alpha)$ which includes $\Nu$ regeneration by $\Tau$ interactions is detailed in~\Cref{app:probs}.

As mentioned above, interaction of $\Nu$  or $\Tau$ with Earth can also lead to a visible shower. The number of events generated by interactions of $\Nu$ is
\begin{equation}
    \frac{\mathrm{d}N_\Nu(\alpha)}{\mathrm{d}\Omega} = \frac{\Phi}{4\pi} \, P^\Nu_\mathrm{exit}(\alpha) \, \sigma \, N_\mathrm{targets} \, \varepsilon \, \Delta t\, ,
\end{equation}
with $N_\mathrm{targets}$ the number of nucleons inside the detector effective volume, $\varepsilon$ the detection efficiency and $P^\Nu_\mathrm{exit}(\alpha)$ the probability for $\Nu$ to arrive to the detector. The full expression of $P^\Nu_\mathrm{exit}(\alpha)$ which includes $\Nu$ regeneration by $\Tau$ interactions is detailed in~\Cref{app:probs}.  The number of events generated by interactions of $\Tau$ is
\begin{equation}
    \frac{\mathrm{d}N_\Tau^\mathrm{int}(\alpha)}{\mathrm{d}\Omega} = \frac{\Phi}{4\pi} \, P^\Tau_\mathrm{exit}(\alpha) \, \sigma \, N_\mathrm{targets} \, \varepsilon \, \Delta t\, .
\end{equation}
We provide further details on the computations and our implementation of the ANITA and IceCube detectors in~\cref{app:probs}.

\subsection{Transient sources \texorpdfstring{\emph{vs}}{vs} diffuse flux\label{sec:steady-or-not}}

There are two scenarios for the incoming particle flux --- $\nu_\tau$ for SM, N for BSM.
It can be transient, i.e., the flux is non-zero only in some time window; or it can be diffuse, i.e., the flux is produced by many sources and is constant in time.

As the IceCube effective area is a factor $\sim 10^2$ smaller than that of ANITA~\cite{ANITA:2021xxh,IceCube:2016uab}, a transient origin for the events cannot be tested by IceCube as long as four transient sources activated \emph{only} in the month that ANITA-IV was flying and \emph{never again} in the nine years of IceCube operation (as the transient rate increases, the flux becomes diffuse).
This admittedly baroque hypothesis would allow even a SM explanation of the anomalous events~\cite{ANITA:2021xxh}.
For the rest of the paper, we focus on the more realistic diffuse flux hypothesis.

\section{ANITA-IV angular self-consistency\label{sec:anita-angular}}

In this section, we show that, despite the small sample size, the observed angular distribution at ANITA provides information on BSM parameters.

\begin{figure}
	\centering
	\includegraphics[width=0.495\linewidth]{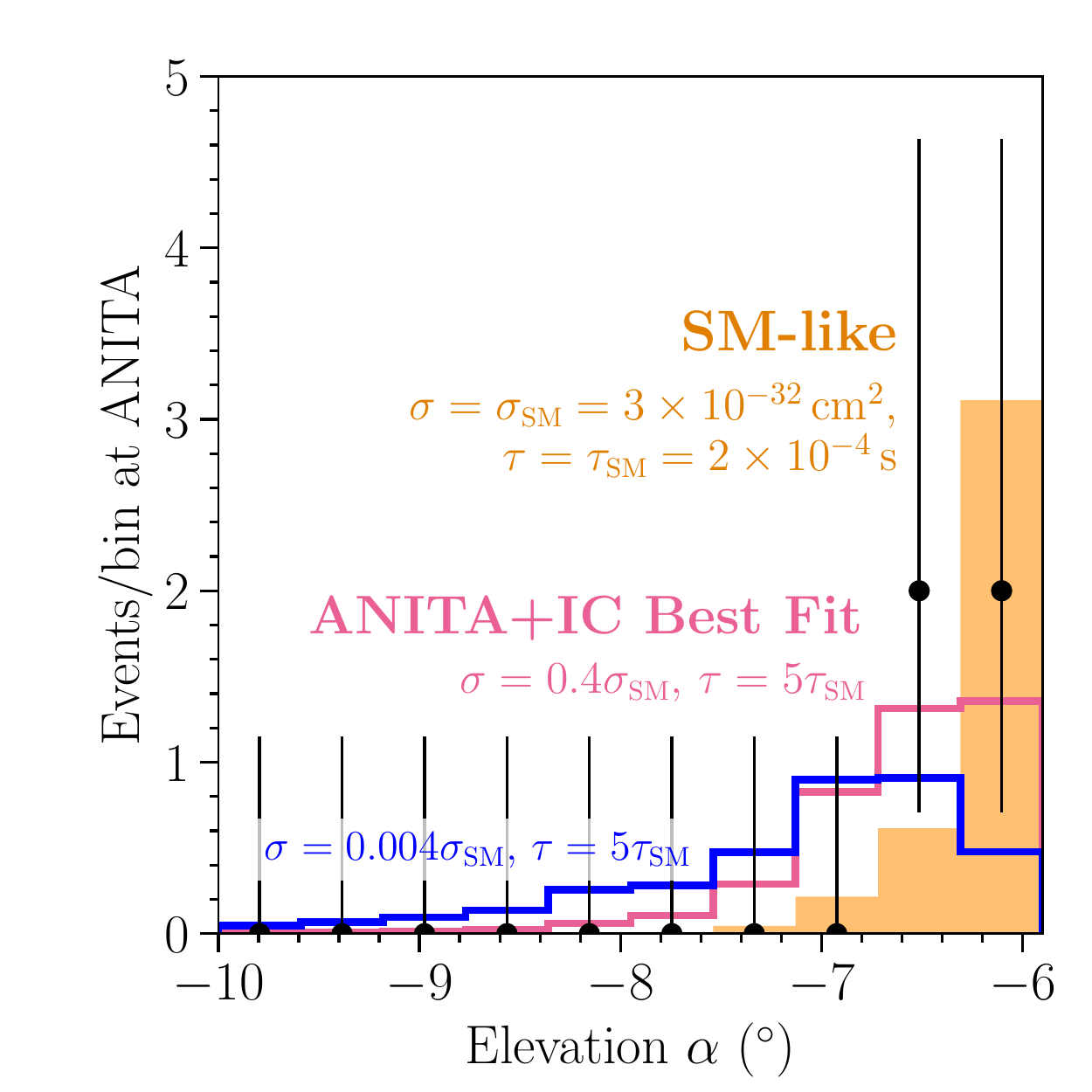}
	\caption{Angular event distribution at ANITA for different values of $\sigma$ and $\tau$ under the diffuse-flux hypothesis. $\sigma$ controls the angular event distribution, whereas $\tau$ mostly controls the normalization (see text). \emph{The incoming directions at ANITA enforce cross sections around the SM value.}}
	\label{fig:anita-histogram}
\end{figure}

\Cref{fig:anita-histogram} shows how the $\Nu$ cross section, $\sigma$, modifies the event distribution. 
Reduced $\sigma$ increases the distance that $\Nu$ can travel inside Earth, making the outgoing $\Tau$ distribution more isotropic. 
Generically, the distribution peaks at angles where the chord length inside Earth equals the mean free path of $\Nu$. 
The distributions are normalized to predict four events at ANITA.

The impact of $\tau$ is less significant.
If $\sigma \gtrsim \sigma_\mathrm{SM}$, the distribution of outgoing T is quite anisotropic, and $\tau$ only controls the probability for them to exit Earth and decay before ANITA, which is degenerate with the overall flux normalization.
Explicitly, very small $\tau$ requires large fluxes because of the suppressed probability to exit Earth, and so do very large $\tau$ because of the suppressed probability to decay before ANITA.
If $\sigma \ll \sigma_\mathrm{SM}$, the distribution of T is more isotropic, and large $\tau$ implies a more isotropic event distribution.

To quantify the agreement with observations, we have performed an unbinned likelihood analysis described in~\Cref{app:ts}.
We include the ANITA detection efficiency and angular resolution, and parametrize the Earth density profile with the Preliminary Earth Reference Model (PREM)~\cite{Dziewonski:1981xy} (our results are robust against different parametrizations). 

\begin{figure}
	\centering
	\includegraphics[width = 0.65\linewidth]{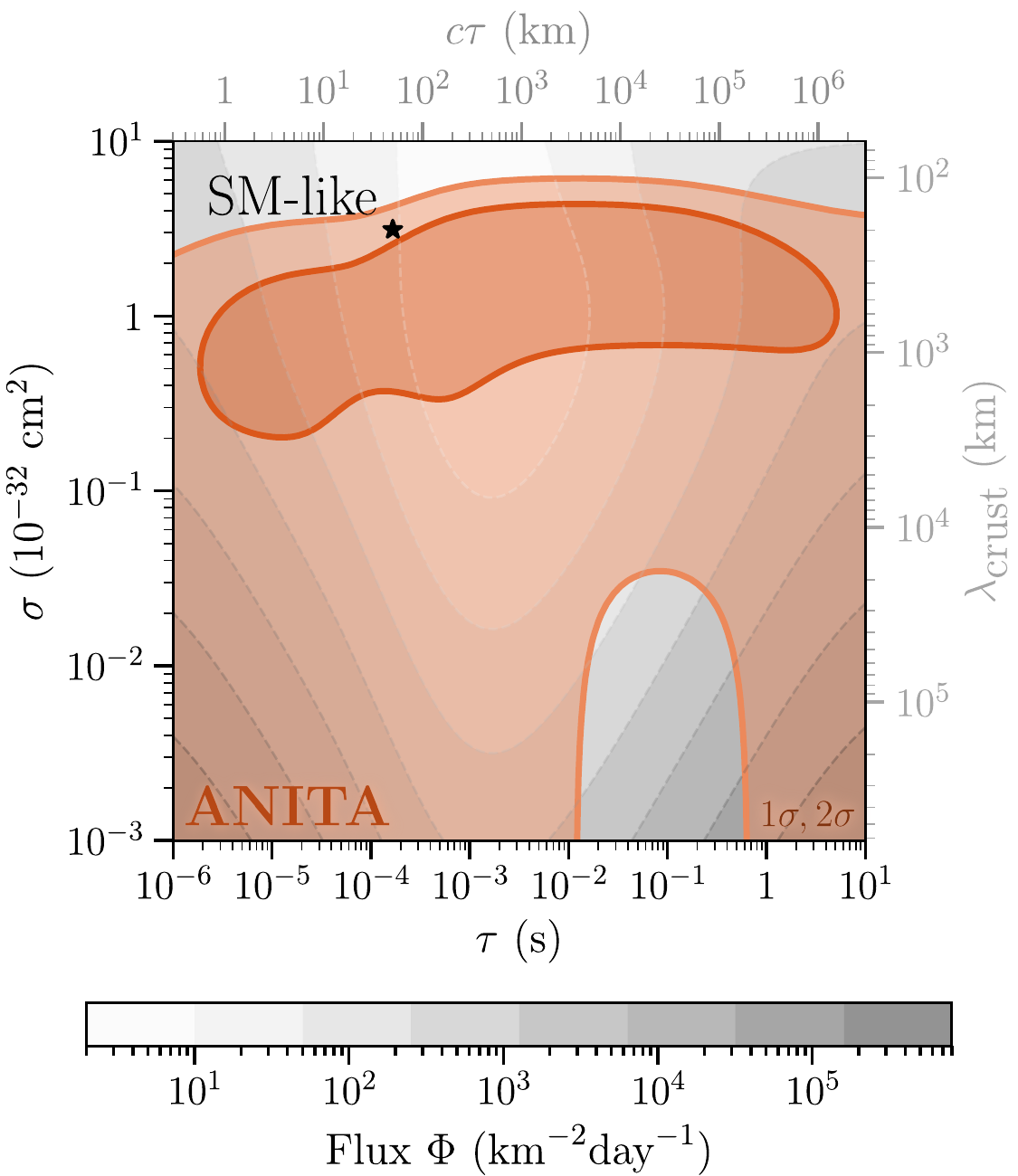}
    \caption{Allowed BSM parameters from ANITA data under the diffuse-flux hypothesis, together with the required incoming N flux. We also show the mean free path of N in the Earth crust, $\lambda_\mathrm{crust}$; and the average distance travelled by the long-lived particle T, $c \tau$.
    }
	\label{fig:ts-ANITA-angulardist}
\end{figure}

\Cref{fig:ts-ANITA-angulardist} shows that the ANITA angular distribution implies a preferred region of BSM parameters.
The star signals the parameters where the BSM sector approximates the SM,
\begin{equation}\label{eq:SMxs}
\begin{split}
    \sigma \sim \sigma_\mathrm{SM} &\simeq 3\times 10^{-32} \, \mathrm{cm}^2 \, , \\ 
	\tau \sim \tau_\mathrm{SM} &\simeq 2\times 10^{-4} \, \mathrm{s} \, .
\end{split}
\end{equation}
Although the approximation is not exact due to the simplifications described in \cref{sec:bsm-expla}, the phenomenology is similar.
We also show the $\Nu$ flux $\Phi$ that would predict four events at ANITA, which is always comparable to or greater than UHE cosmic-rays fluxes ($\sim 1 \, \mathrm{km}^{-2} \mathrm{day}^{-1}$ at 1 EeV~\cite{PierreAuger:2021hun}).
As described above, the flux is strongly correlated with $\tau$; slightly increasing $\tau$ with respect to $\tau_\mathrm{SM}$ reduces the required flux by an order of magnitude.

ANITA data excludes large $\sigma$, because the events would peak closer to the horizon.
For small $\sigma$, $10^{-2}\, \mathrm{s}\lesssim \tau\lesssim 1\, \mathrm{s}$ is excluded because the event distribution would be too isotropic.
$\tau \gtrsim 1 \, \mathrm{s}$ is allowed for small $\sigma$ because most $\Tau$ decay after ANITA, and the events are produced by interactions of $\Nu$ with the atmosphere at the expense of very large fluxes.

We conclude that a large region of parameter space in our model independent framework, including the SM-like scenario, is consistent with ANITA data.
Below, we show that this is challenged by null observations from IceCube.

\section{Interplay with IceCube\label{sec:icecube-interplay}}

The IceCube experiment is sensitive to the same flux that produces events in ANITA~\cite{Safa:2019ege,IceCube:2020gbx,Arguelles:2022aum}.
Although IceCube has a smaller effective volume, its larger angular aperture and observation time allow to test the origin of the ANITA events.
In this section, we describe such a test, pointing out the parameter region where both experiments could be compatible.

\begin{figure}
	\centering
	\includegraphics[width=0.495\linewidth]{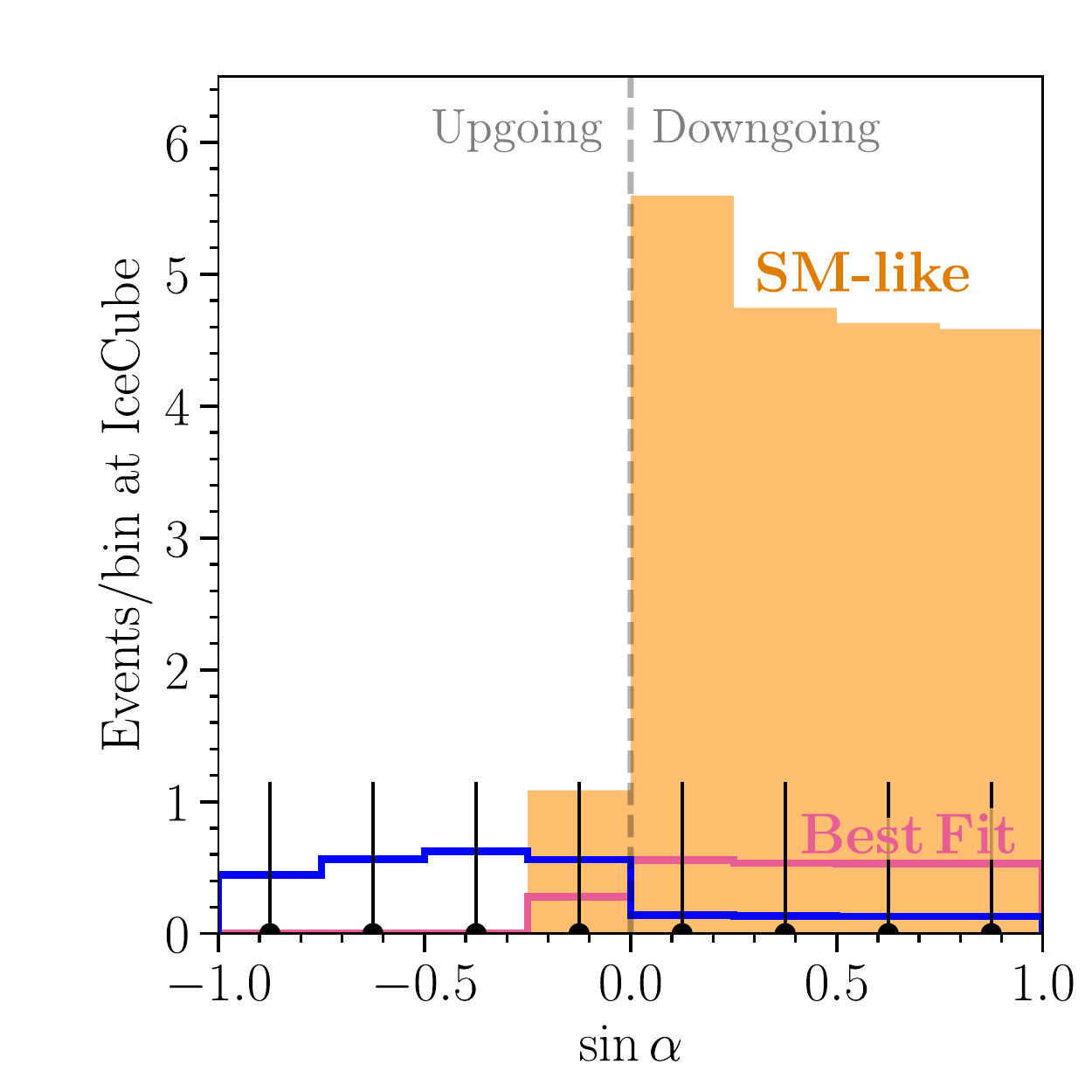}
	\caption{Same as \cref{fig:anita-histogram}, but for IceCube. The normalizations are fixed to predict 4 events at ANITA. \emph{IceCube is mostly sensitive to the flux normalization required to explain 4 events in ANITA}.}
	\label{fig:ic-histogram}
\end{figure}

\Cref{fig:ic-histogram} shows that explanations of the ANITA events are challenged by the lack of observations at IceCube~\cite{IceCube:2018fhm}.
We show the expected event distribution at IceCube, for the same $\sigma$ and $\tau$ as~\cref{fig:anita-histogram} and the flux normalization that predicts four events at ANITA.

Most of the events predicted at IceCube are due to interactions of N with Earth (as we assume that such interactions generate showers), and for SM-like or larger cross sections, they are dominantly downgoing because of Earth attenuation.
The sensitivity of IceCube to $\tau$ is indirect but key for its compatibility with ANITA's measurements.
As described above, slightly increasing $\tau$ with respect to $\tau_\mathrm{SM}$ reduces the required flux, and the predictions are then compatible with null observations at IceCube.
For extremely low values of $\sigma$ (blue histogram), the contribution from N interactions in IceCube is suppressed, and most of the events are generated by upgoing T-s produced by interactions of N with Earth.

\begin{figure}
	\centering
	\includegraphics[width = 0.495\linewidth]{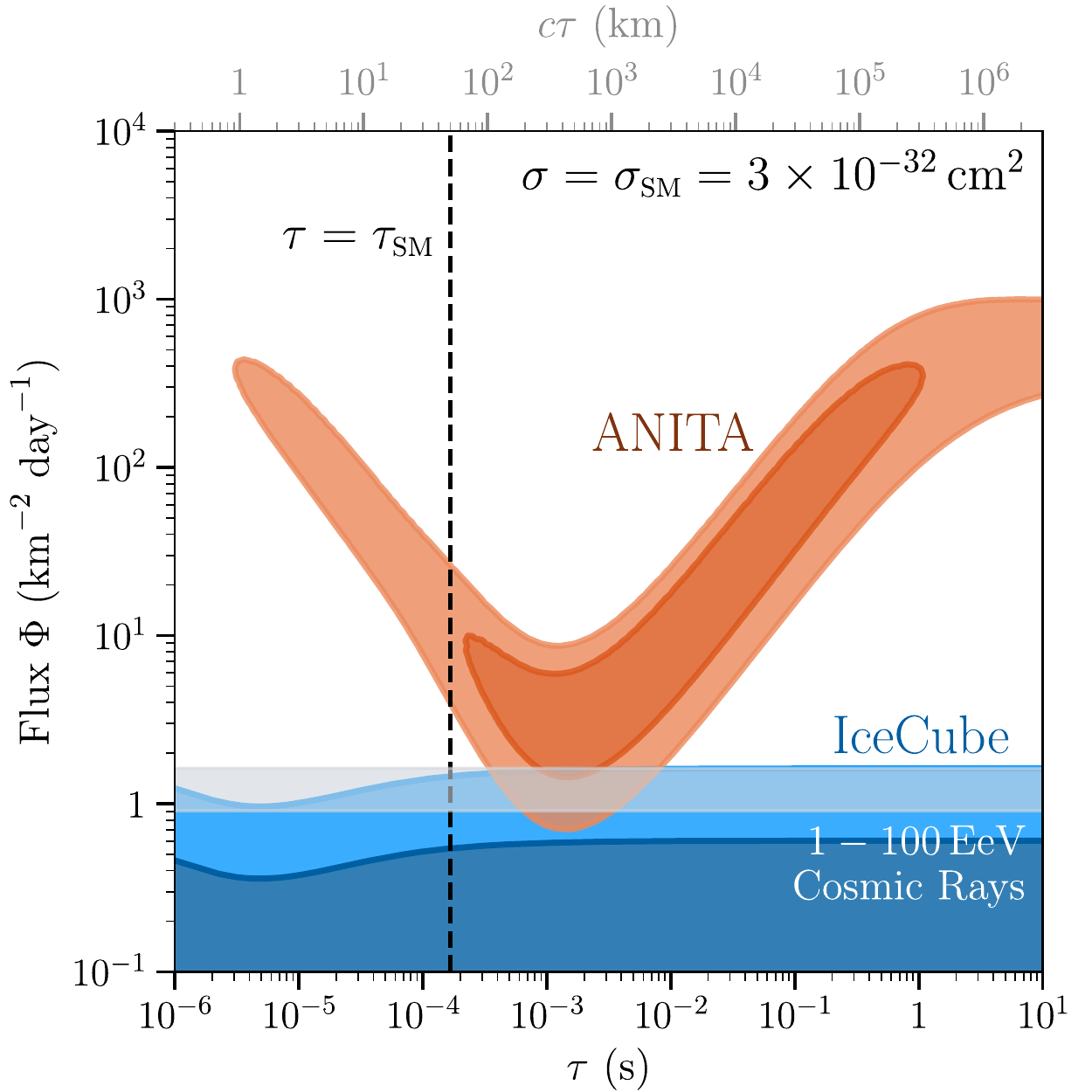}
	\includegraphics[width = 0.495\linewidth]{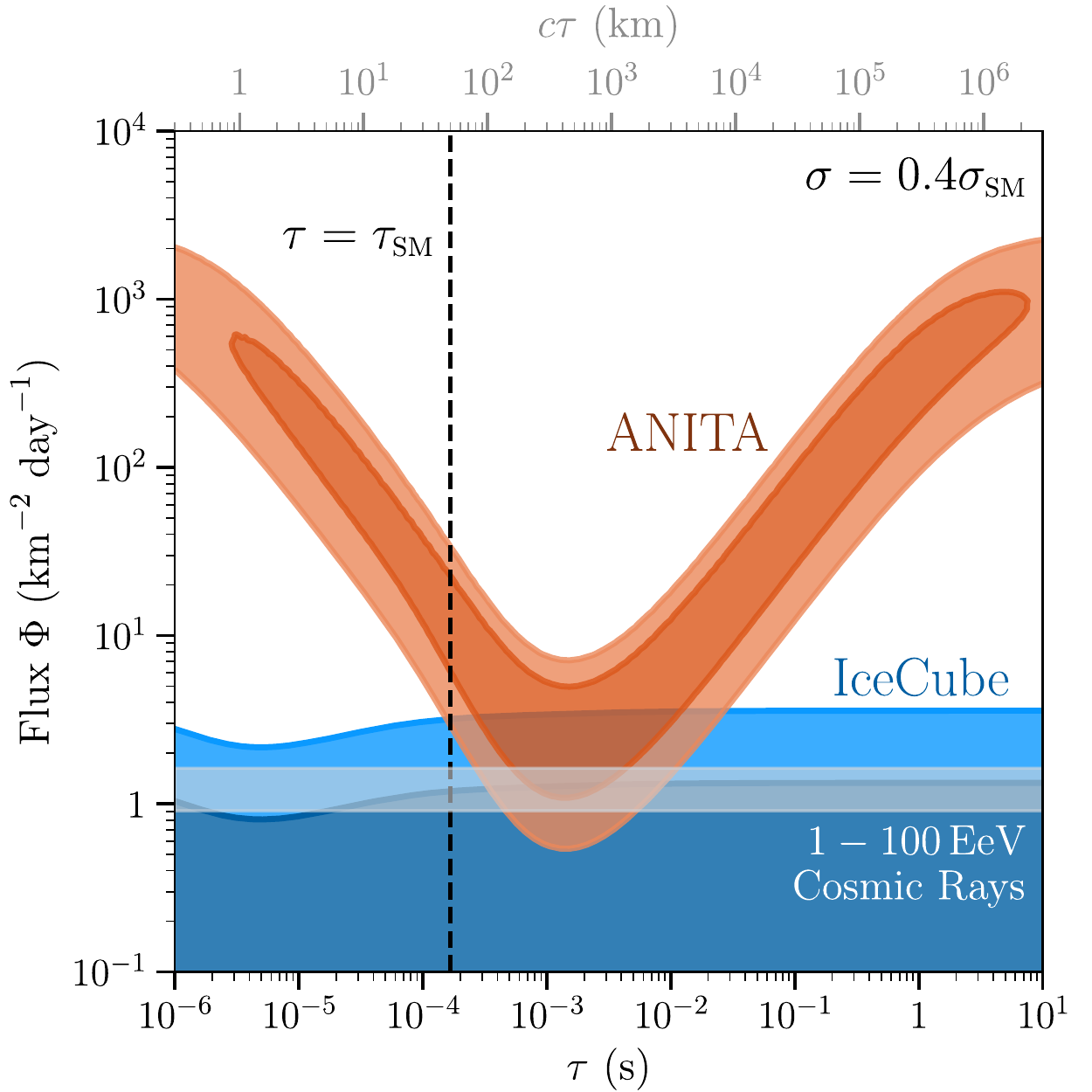}
 	\caption{Allowed regions for $\tau$ and the flux normalization $\Phi$ from ANITA and IceCube independently. We show for orientation the UHE Cosmic Ray flux~\cite{PierreAuger:2021hun}. \emph{$\tau$ and $\Phi$ are strongly correlated in ANITA, while IceCube is insensitive to $\tau$. The BSM parameters can partially alleviate the tension.}}
	\label{fig:ts-ANITA-IC-normalizations}
\end{figure}

\Cref{fig:ts-ANITA-IC-normalizations} shows how the correlation between lifetime and flux affects the interplay between ANITA and IceCube, with only some values of $\tau$ allowing $\Phi$ consistent both with the observations at ANITA and the lack of events at IceCube. 
We show the allowed values of $\tau$ and $\Phi$ for fixed $\sigma$, obtained using an unbinned likelihood described in \cref{app:ts}. For ANITA, $\Phi$ is always comparable to or higher than the cosmic-ray flux as mentioned above.
As the figure shows, IceCube is mostly sensitive to interactions of N, i.e., to its flux $\Phi$.
In turn, ANITA mostly detects T decays, and $\Phi$ is very correlated with $\tau$ because T must exit Earth \emph{and} decay before ANITA:  small $\tau$ spoil the former and large $\tau$ spoil the latter.

\begin{figure}
	\centering
	\includegraphics[width = 0.65\linewidth]{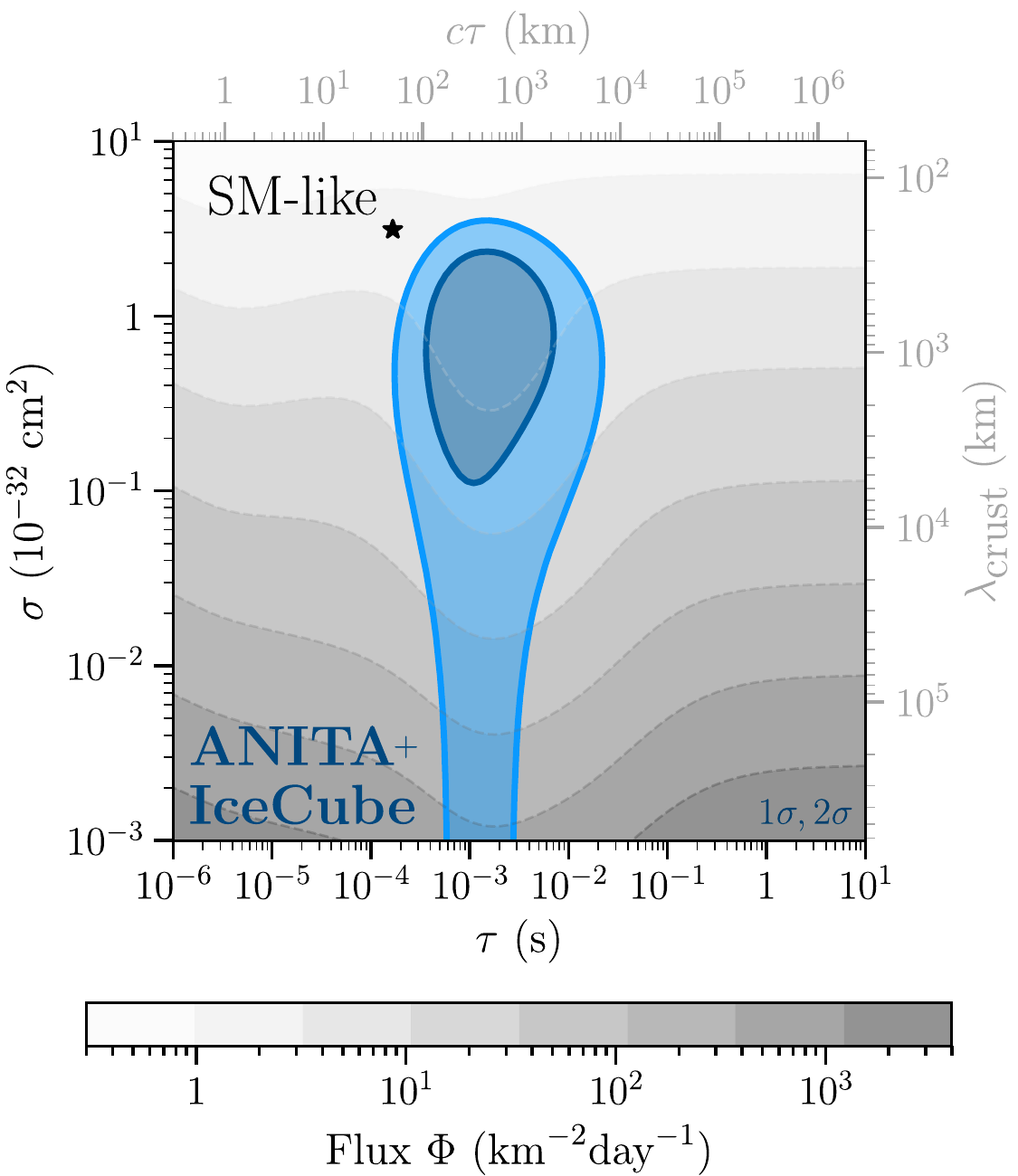}
    \caption{Allowed BSM parameters the combination of ANITA and IceCube (blue) under the diffuse-flux hypothesis, together with the required incoming particle flux. The best fit predicts 1.2 events after 9 years of IceCube operation. \emph{Although the SM-like scenario is excluded at $\sim 3\sigma$, BSM models may explain the ANITA anomalous events, and signals would be observable in IceCube-Gen2}.}
	\label{fig:ts-ANITA-IC-total}
\end{figure}

\Cref{fig:ts-ANITA-IC-total} shows the ranges of $\sigma$ and $\tau$ allowed by a combined analysis of ANITA and IceCube, together with the best-fit flux $\Phi$.
From \cref{fig:ts-ANITA-angulardist}, non-observation of events at IceCube only allows regions with small $\Phi$, which implies $\tau \sim 10^{-3} \, \mathrm{s}$.
There is also some information on $\sigma$ because it controls both the angular distribution at ANITA and the number of events at IceCube due to N interactions.
Altogether, this implies closed allowed regions up to ${\sim 2\sigma}$ (we recall that the significance of the anomalous events is $\sim 3\sigma$~\cite{ANITA:2020gmv, ANITA:2021xxh}), with the best-fit point at $\sigma = 8.9 \times 10^{-33}\,\text{cm}^2$, $\tau = 1.3 \times 10^{-3}\,\textrm{s}$, and $\Phi = 1.8 \, \mathrm{km}^{-2}\, \mathrm{day}^{-1}$.
The best-fit parameters predict 1.2 events at IceCube after nine years of operation, and the best-fit flux at every point in the parameter space predicts at least 0.9 events.

Overall, a BSM interpretation relaxes the tension between the ANITA anomalous events and IceCube, but does not fully remove it.
The combined allowed regions always predict $\mathcal{O}(1)$ events at IceCube, so if the anomalous events have a particle physics origin parametrized by our set of BSM models, the signal should be observable in the future.

\section{Particle physics models\label{sec:models}}
Above, we have demonstrated in a model independent approach that BSM scenarios could accommodate the ANITA observations and the absence of any signal in IceCube. In this section, we discuss explicit particle physics models that can provide the required ingredients.

As our best fit is not very far from the SM-like prediction, an appealing possibility is to consider BSM models related to neutrinos, such as scenarios involving heavy neutral leptons that mix with the SM neutrinos and further couple to other fields belonging to a richer Dark Sector. In such scenarios, the SM neutrinos could play the role of $\Nu$.

However, one of the main issues in searching for a successful particle physics model is the origin of the flux of $\Nu$ particles. Current neutrino flux limits are $\Phi \lesssim 0.1 \, \mathrm{km^{-2}\,day^{-1}}$~\cite{IceCube:2018fhm, PierreAuger:2019ens}, well below the required $\Nu$ flux (see \cref{fig:ts-ANITA-IC-total}). Thus, a more exotic primary particle is generically required. A Dark Matter (DM) origin is probably the less exotic possibility. Such option has been recently put forward~\cite{Heurtier:2019git,Hooper:2019ytr,Cline:2019snp,Heurtier:2019rkz} to explain the two upgoing highly anomalous events previously observed by ANITA~\cite{ANITA:2016vrp,ANITA:2018sgj}. The decay of extremely heavy DM (in the EeV range) to $\Nu$ can produce very large fluxes~\cite{Cirelli:2010xx}
\begin{equation}
    \Phi_\mathrm{N} = \frac{1}{\tau_\mathrm{DM} m_\mathrm{DM}} \int \rho_\mathrm{DM}\,\mathrm{d}s\, ,
\end{equation}
where $\tau_\mathrm{DM} > 10^{19}\, \mathrm{s}$~\cite{Alvi:2022aam} is the DM lifetime and $m_\mathrm{DM}$ its mass, and the integral is along the line of sight $s$. Numerically, $\int \rho_\mathrm{DM}\,\mathrm{d}s \sim 5 \times 10^{22} \, \mathrm{GeV/cm^2}$~\cite{Cirelli:2010xx}. If $m_\mathrm{DM} = 1 \, \mathrm{EeV}$, then $\Phi_\mathrm{N}$ can be as large as $10^{12} \, \mathrm{km^{-2} \, day^{-1}}$, well in agreement with \cref{fig:ts-ANITA-IC-total}.  

Following this idea, we consider a toy model involving a dark sector that includes an extremely heavy DM candidate and two dark fermions playing the role of our $\Nu$ and $\Tau$ particles. Adding an extra $U_X(1)$ gauge symmetry in the Dark Sector, we introduce a new dark boson $X_\mu$ that mixes with the SM photon via kinetic mixing~\cite{Holdom:1985ag,Okun:1982xi}, allowing the dark fermions to interact with ordinary matter and have nonzero $\sigma$ and $\tau$. 

In more detail, the DM can be a scalar SM singlet $\phi$ that decays via a Yukawa coupling to a stable fermion $\chi_1$, a SM singlet that would play the role of N. The role of T would be played by a second fermion singlet $\chi_2$ (heavier than $\chi_1$) which, choosing the dark charges of both fermions appropriately, couples to $X_\mu$ just via a $g_D\bar{\chi}_2\gamma^\mu\chi_1X_\mu$ term in the Lagrangian. This way, $\chi_1$ interacts with Earth nuclei via kinetic mixing generating a $\chi_2$, which subsequently decays via $\chi_2\rightarrow \chi_1 +\rm{shower}$, the shower being originated through kinetic mixing with the ordinary photon. The same model was proposed to fit the previous ANITA anomalous events~\cite{Heurtier:2019rkz}. Following Ref.~\cite{Heurtier:2019rkz}, we find that for $g_D=1.1$, particle masses $m_X=1.5$ GeV, $m_{\chi_2}=0.7$ GeV, $m_{\chi_1}=0.5$ GeV, and kinetic mixing $\epsilon = 7\cdot 10^{-3}$, the values of $\sigma$ and $\tau$ are in the right ballpark in agreement with \cref{fig:ts-ANITA-IC-total} (these are different from the values considered in Ref.~\cite{Heurtier:2019rkz} to explain the previous anomalous events). These parameter values are currently allowed (see, e.g., Fig.~5.1 in Ref.~\cite{Mongillo:2023hbs}). The constraints on this scenario are weaker than for minimal dark photon scenarios --- where the dark photon decays predominantly either into SM visible particles (visible dark photons) or into missing energy (invisible dark photons) --- because $X_\mu$ decays involve both SM and dark particles (\emph{semi-visible} dark photons).

The proposal considered here is an extension of the so-called inelastic DM models~\cite{Tucker-Smith:2001myb}. In such models $\chi_1$ constitutes the thermal DM relic, and there are typically no new scalars such as our singlet $\phi$. The $g_\mu-2$ anomaly can also be accommodated~\cite{Mohlabeng:2019vrz}, but it has been shown recently that both phenomena cannot be simultaneously explained in most part of the parameter space~\cite{Mongillo:2023hbs} (see also a review of semi-visible light dark photon models in Ref.~\cite{Abdullahi:2023tyk}). Instead, in the model considered here the DM is mainly composed of a super-heavy scalar singlet $\phi$, while $\chi_1$ is a subdominant component. This super-heavy DM can be generated for instance via freeze-in~\cite{Kolb:2017jvz} or at the end of inflation~\cite{Greene:1997ge,Chung:1998ua,Chung:1998rq,Kannike:2016jfs}.

\section{Summary and conclusions\label{sec:conclu}}

The fourth flight of ANITA found four upgoing UHE events coming from about a degree below the horizon. Explaining these events within the SM implies a flux of $\nu_{\tau}$ inconsistent with the non-observations in Auger or IceCube. Here, we have performed a model-independent analysis of IceCube and ANITA-IV results for BSM scenarios. We consider a conservative approach assuming an incoming flux of generic particles $\Nu$ only at the energy window where ANITA has observed the anomalous events. The $\Nu$ particles interact with Earth nucleons --- with cross section $\sigma$ --- generating primary showers along with long lived particles $T$ that decay --- with lifetime $\tau$ --- producing another shower (see \cref{fig:bsm}). Such formalism can be applied to better understand the origin of any future observation.

Therefore, the set of models under consideration are determined by three parameters: the flux ($\Phi$) of the incoming $\Nu$ particles, their cross section with nucleons ($\sigma$), and the lifetime ($\tau$) of the secondary long-lived T particles produced after the interaction. Performing a statistical analysis of the four events observed by ANITA, we found that the low statistics allows large ranges for $\sigma$ and $\tau$ compatible with a SM-like explanation of the signals. The angular distribution of the events excludes $\sigma \gtrsim 5 \times 10^{-32}\,\mathrm{cm}^2$, where the events would peak closer to the horizon, together with $\sigma \lesssim 5 \times 10^{-34}\,\mathrm{cm}^2$ and $10^{-2} \lesssim \tau \lesssim 1$, where the distribution would be too isotropic.

The large observation time and its large angular acceptance makes IceCube an excellent candidate to explore the flux observed by ANITA. The main sensitivity of IceCube to this set of models comes from interactions of N with Earth. Null results in IceCube are compatible with ANITA for lifetimes $\tau \sim 10^{-3}$~s and cross sections $\sigma \lesssim 3 \times 10^{-32}\,\mathrm{cm}^2$ at $2\sigma$. 

Finally, we provide a concrete scenario based on super-heavy scalar Dark Matter that can explain ANITA-IV and IceCube data. The decay of the Dark Matter into a stable dark fermion $\chi_{1}$, via a Yukawa coupling, can account for the large fluxes needed ($\Phi \gtrsim 1~\text{km}^{-2}\text{day}^{-1}$). In the scenario considered in this work, the Dark Sector is also enlarged by an unstable dark fermion $\chi_{2}$, heavier than $\chi_{1}$, which plays the role of $T$. Adding, an extra $U_{X}(1)$ dark gauge symmetry, we also introduce a dark boson with a non diagonal coupling to the dark fermions and kinetic mixing with the SM photon. This allows $\chi_{1}$ to interact with Earth nucleons, generating a $\chi_2$ which subsequently decays via $\chi_2\rightarrow \chi_1+\rm{shower}$ with the shower generated through kinetic mixing.
Previously proposed as an explanation of the anomalous events observed by ANITA in previous flights~\cite{Heurtier:2019rkz}, this scenario can account for the new anomalous events with large kinetic mixings ($\sim 10^{-2}$), and masses around the GeV scale ($m_{X}\sim 2$~GeV, $m_{\chi_2}\sim 1$ GeV, $m_{\chi_1}\sim 0.5$ GeV). Further model-building work can explore other options.

The ANITA experiment is starting to probe uncharted land. First results are already anomalous, and understanding their BSM or background origin is pressing to ensure the success of future, more ambitious, UHE neutrino detectors. As we have shown, BSM explanations predict that anomalous events should be observed in other experiments too. The upcoming flight of PUEO~\cite{PUEO:2020bnn}, with a detection principle similar to ANITA, and the future IceCube-Gen2 upgrade~\cite{IceCube-Gen2:2020qha} will thus be unique opportunities to gain insight into the potential signals and backgrounds of the UHE landscape.

\section*{Acknowledgements}

CAA and IMS are supported by the Faculty of Arts and Sciences of Harvard University.
Additionally, CAA and IMS are supported by the Alfred P. Sloan Foundation. 
This work has received partial support from the European Union's Horizon 2020 research and innovation programme under the Marie Sk\l odowska-Curie grant agreement No 860881-HIDDeN and the Marie Skłodowska-Curie Staff Exchange  grant agreement No 101086085 – ASYMMETRY. 
TB and JS acknowledge financial support from the Spanish grants PID2019-108122GBC32, PID2019-105614GB-C21, and from the State Agency for Research of the Spanish Ministry of Science and Innovation through the ``Unit of Excellence María de Maeztu 2020-2023'' award to the Institute of Cosmos Sciences (CEX2019-000918-M). JLP acknowledges support from Generalitat Valenciana through the plan GenT program (CIDEGENT/2018/019) and from the Spanish Ministerio de Ciencia e Innovacion through the project PID2020-113644GB-I00.

\bibliography{bib}

\clearpage

\appendix

\section{Impact of T absorption}\label{app:absorption}
In the main text, we assume that $\Tau$ interacts with Earth with the same cross section that produces it, $\sigma$. This is to be expected in the simplest BSM extensions. However, the $\Nu$ absorption cross section $\sigma_\Nu$ and the $\Tau$ absorption cross section $\sigma_\Tau$ could in principle be different. In particular, $\sigma_\Tau=0$ would maximize the compatibility between ANITA and IceCube. 
The number of events in ANITA would increase due to fewer $\Tau$ particles being absorbed by the Earth, and the number of events in IceCube would slightly decrease due to fewer interaction of T particles with the detector.

\begin{figure}[hbtp]
	\centering
	\includegraphics[width = 0.495\linewidth]{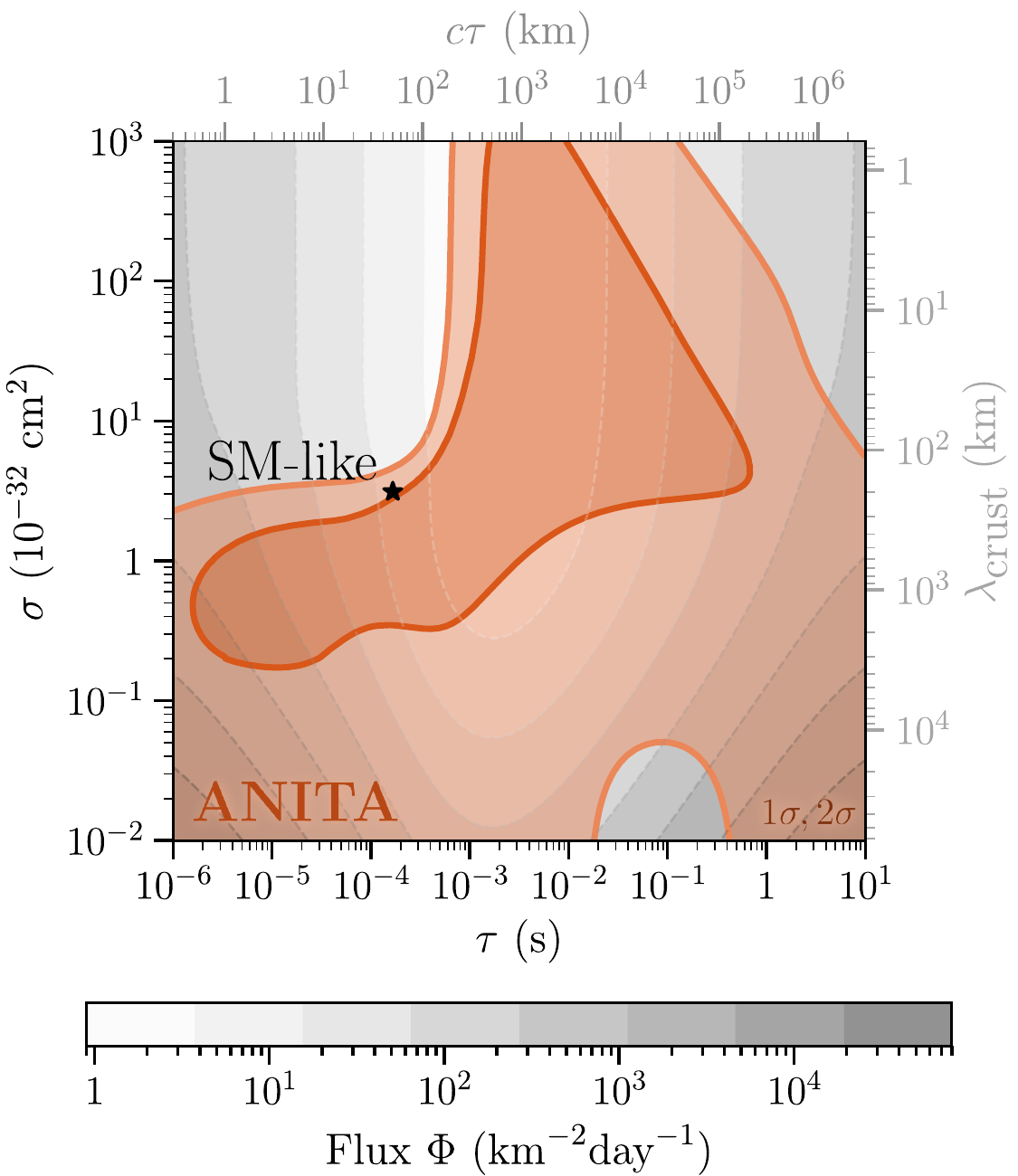}
	\includegraphics[width = 0.495\linewidth]{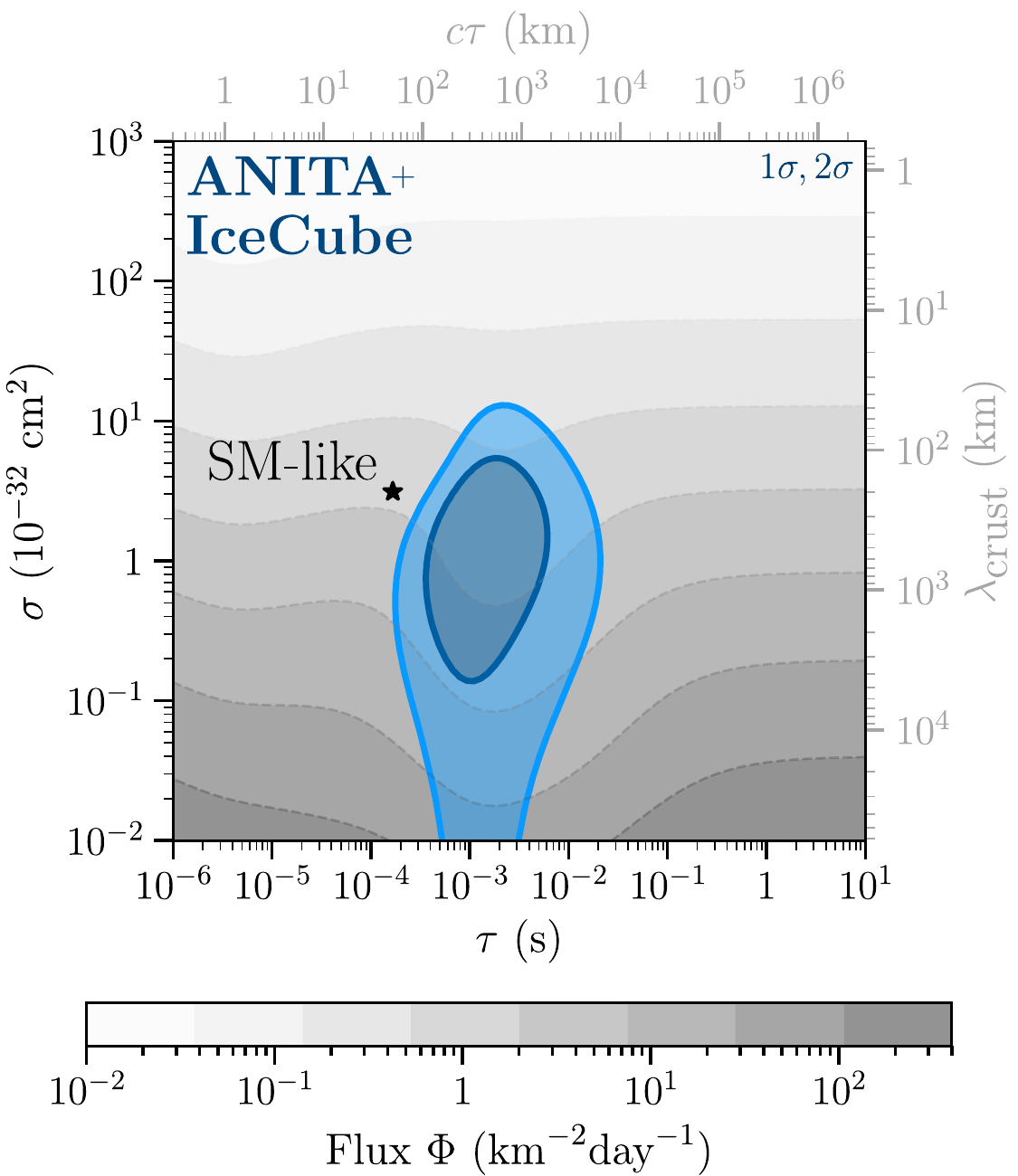}
    \caption{Allowed BSM parameters from ANITA (left, red) and the combination of ANITA and IceCube (right, blue) under the diffuse-flux hypothesis, together with the required incoming particle flux, if we remove interactions of T with Earth. Note that, by neglecting T absorption, the ``SM-like'' point resembles less the SM than the choice in the main text. \emph{ANITA alone is sensitive to $\Tau$ interaction with matter, but its combination with IceCube is mostly not.}}
	\label{fig:ts-total-noabsorption}
\end{figure}

\Cref{fig:ts-total-noabsorption} shows the results obtained following the same procedure as in the main text but setting  $\sigma_\Tau = 0$. The left panel shows the results including only ANITA-IV data. There are some qualitative differences with respect to the $\sigma_\Nu=\sigma_\Tau$ case in \cref{fig:ts-ANITA-angulardist}. For $10^{-3}\,\mathrm{s}\lesssim \tau\lesssim 1\,\mathrm{s}$, larger $\sigma$ is allowed because, even though N is always absorbed, $\Tau$ can exit Earth and generate the anomalous events. However, data disfavors too long lifetimes because the event distribution would be too isotropic. 

In the right panel we show the results with both ANITA and IceCube data. 
The combination leads to an allowed region similar to the one shown in~\cref{fig:ts-ANITA-IC-total} for the scenario with $\sigma_\Tau = \sigma_N$. This is because the large-$\sigma_\Nu$ region that is allowed by ANITA if $\sigma_\Tau = 0$ would produce too many downgoing events in IceCube. We thus conclude that the conclusions drawn in the main text are robust against assuming that $\Tau$ does not interact with Earth.

\section{Details on the computations}\label{app:probs}
In this appendix, we provide details on our event simulation. We describe the geometry of the trajectories, the computation of the number of events, and the details of particle propagation inside Earth.

\subsection{Geometry}
For ANITA, the elevation angle $\alpha$ can be related to the zenith angle $\theta$ of the particle T at its exit point by 
\begin{equation}\label{eq:angle-conversion}
	\sin\theta = \cos\alpha \left(1+\frac{h_{\text{ant}}}{R_\oplus}\right)\, ,
\end{equation}
with $h_\mathrm{ant}$ the height of the ANITA antenna and $R_\oplus$ the Earth radius. However, radio waves are refracted during propagation, which modifies the relation between $\theta$ and the observed elevation. We implement this by effectively increasing the Earth radius by a fudge factor $1.13$ that reproduces the horizon elevation angle as a function of $h_\text{ant}$ as provided by the ANITA collaboration (see Table 1 in Ref.~\cite{ANITA:2020gmv}).

For the IceCube experiment, the relation is
\begin{equation}
    \sin\theta = -\cos\alpha\left(1-\frac{h_{\textrm{exp}}}{R_\oplus}\right)\, ,
\end{equation}
with $h_{\text{exp}} = 1\, \mathrm{km}$ the depth of IceCube.

\subsection{Number of events}
We parametrize the incoming flux by a normalization constant, the energy spectrum and the angular distribution,
\begin{equation}
	\Phi(\Omega, E) = \Phi_0\, f_E(E)\, f_\Omega(\Omega)\, ,
\end{equation}where $E$ is energy, $\Omega$ the solid angle ($\mathrm{d}\Omega = \sin\theta\, \mathrm{d}\theta\, \mathrm{d}\varphi$), and $\Phi_0$ the flux per unit area, time, solid angle, and energy. The expected number of events per unit solid angle and energy is
\begin{equation}
     \frac{\mathrm{d}N(\theta)}{\mathrm{d}\Omega\,\mathrm{d}E} = \Phi_0\, \Delta t\, f_E(E)\, f_\Omega(\Omega)\, \mathcal{A}(\Omega, E)\, .
\end{equation}Here $\Delta t$ is the total observation time; and $\mathcal{A}$ includes the geometric area of the detector, detection efficiency, the absorption of the flux inside Earth, and the probability for the flux to produce a detectable signal. It encodes all the details of the propagation and absorption models, as we explain next.  We assume $\mathcal{A}$ has axial symmetry and does not depend strongly on energy~\cite{ANITA:2021xxh,IceCube:2016uab}. Then, $\mathcal{A}(\Omega, E) = \mathcal{A}(\theta)$.

We are interested in the expected number of events per solid angle $\Omega$, so we integrate over energy
\begin{equation}\label{eq:exp_events_0}
     \frac{\mathrm{d}N(\theta)}{\mathrm{d}\Omega} = 
     \left(\Phi_0\int_{E_{\text{min}}}^{E_\text{max}} \mathrm{d}E\,  f_E(E)\right) f_\Omega(\Omega) \,  \Delta t \, \mathcal{A}(\theta)   \equiv 
     \Phi\, f_\Omega(\Omega)\, \Delta t \, \mathcal{A}(\theta)
     \, .
\end{equation}
Experiments do not perfectly reconstruct the true angle of the incoming particle, $\theta^{\text{true}}$. To take this into account, we assume Gaussian angular uncertainty $\Delta\theta$. For ANITA-IV, $\Delta\theta$ is reported in Ref.~\cite{ANITA:2020gmv}. Then, the expected number of events per unit solid angle as a function of the reconstructed angle $\theta^{\text{rec}}$ is given by
\begin{equation}
\begin{split}
    \frac{\mathrm{d}\bar{N}(\theta^{\text{rec}})}{\mathrm{d}\Omega} &= 
	\int \mathrm{d}\theta^{\text{true}} \frac{1}{\sqrt{2\pi}\Delta\theta} 
    \exp\left[ -\frac{(\theta^{\text{rec}}-\theta^{\text{true}})^2}{2(\Delta\theta)^2} \right] 
    \frac{\mathrm{d}N(\theta^{\text{true}})}{\mathrm{d}\Omega} = \\ &=
	\frac{\Phi \Delta t}{\sqrt{2\pi}\Delta\theta} \int \mathrm{d}\theta^{\text{true}} f_\Omega(\theta^{\text{true}},\varphi) \mathcal{A}(\theta^{\text{true}})   \exp\left[-\frac{(\theta^{\text{rec}}-\theta^{\text{true}})^2}{2(\Delta\theta)^2} \right] \, .
\end{split}
\end{equation}

\subsection{Absorption and detection processes}
In our model, events come either from $\Tau$ decays/interactions or from $\Nu$ interactions. We separate these contributions as $\mathcal{A} = \mathcal{A}_{\Tau}^{\mathrm{dec}} + \mathcal{A}_{\Tau}^{\mathrm{int}}  + \mathcal{A}_{\Nu}$. On the one hand, 
\begin{equation}
    \mathcal{A}_{\Tau}^{\mathrm{dec}}(\theta) = P^\Tau_\mathrm{exit}(\theta) \, P^\Tau_\mathrm{decay}(\theta) \, A_\mathrm{eff}(\theta)\, .
\end{equation} Here $A_\mathrm{eff}(\theta)$ is the area to which the detector is sensitive, including detection efficiency. For ANITA, the geometric area and trigger efficiency have been extracted from Figure 10 in Ref.~\cite{ANITA:2021xxh}. For IceCube, we have set the geometric area to $1\, \mathrm{km}^2$ and the detection efficiency has been extracted from Ref.~\cite{Palomares-Ruiz:2015mka}. Then, 
$P_{\text{decay}}^\Tau(\theta) \equiv 1 - e^{-d(\theta)/\tau}$ is the probability for $\Tau$ to decay inside the effective volume of length $d(\theta)$. For ANITA, $d(\theta)$ is the distance between the exit point and the radio antenna, given by
\begin{equation}
	d(\theta) = - R_\oplus\frac{\cos(\theta-\alpha(\theta))}{\cos\alpha(\theta)}\sim \mathcal{O}(500\textrm{ km}) \, .
\end{equation}For IceCube, we set $d(\theta) = 1\textrm{ km}$. Finally, $P^\Tau_\mathrm{exit}(\theta)$ is the probability for $\Tau$ to be produced and arrive to the effective volume, that we compute below. Altogether, the total number of expected events from $\Tau$ decay per unit solid angle is
\begin{equation}
    \frac{\mathrm{d}N_\Tau^\mathrm{dec}(\theta)}{\mathrm{d}\Omega} = 
    \Phi f_\Omega(\Omega) \, \mathcal{A}_\Tau^\mathrm{dec}(\theta) \, \Delta t  = 
    \Phi f_\Omega(\Omega) \, P^\Tau_\mathrm{exit}(\theta) \, P^\Tau_\mathrm{decay}(\theta) \, A_\mathrm{eff}(\theta) \, \Delta t \, .
\end{equation}On the other hand, the contribution from $\Tau$ and $\Nu$ interactions is
\begin{align}
    \mathcal{A}_\Tau^\mathrm{int}(\theta) &=  P^\Tau_\mathrm{exit}(\theta)\, N_\mathrm{targets} \, \sigma \, \varepsilon\, \, , \\
    \mathcal{A}_\Nu(\theta) &=  P^\Nu_\mathrm{exit}(\theta)\, N_\mathrm{targets} \, \sigma \, \varepsilon\, ,
\end{align}respectively. Here, $N_\mathrm{targets}$ is the number of targets in the detector, $\varepsilon$ the detection efficiency, and $P^\Nu_\mathrm{exit}(\theta)$ is the probability for an $\Nu$ particle to arrive to the effective volume that we compute below.

The number of expected events from $\Tau$ and $\Nu$ interactions per unit of solid angle is
\begin{align}
    \frac{\mathrm{d}N_\Tau^\mathrm{int}(\theta)}{\mathrm{d}\Omega} &= 
    \Phi f_\Omega(\Omega) \, \mathcal{A}_\Tau^{\mathrm{int}}(\theta) \, \Delta t  = 
    \Phi f_\Omega(\Omega) \, P^\Tau_\mathrm{exit}(\theta)\, N_\mathrm{targets} \, \sigma \, \varepsilon\, \Delta t \, , \\
    \frac{\mathrm{d}N_\Nu(\theta)}{\mathrm{d}\Omega} &= 
    \Phi f_\Omega(\Omega) \, \mathcal{A}_\Nu(\theta) \, \Delta t  = 
    \Phi f_\Omega(\Omega) \, P^\Nu_\mathrm{exit}(\theta)\, N_\mathrm{targets} \, \sigma \, \varepsilon\, \Delta t \, ,
\end{align}respectively. The total number of events is $\frac{\mathrm{d}N(\theta)}{\mathrm{d}\Omega} = \frac{\mathrm{d}N_\Tau^\mathrm{dec}(\theta)}{\mathrm{d}\Omega} + \frac{\mathrm{d}N_\Tau^\mathrm{int}(\theta)}{\mathrm{d}\Omega}+ \frac{\mathrm{d}N_\Nu(\theta)}{\mathrm{d}\Omega}$. 

\subsection{N exit probability}
The probabilities defined in the previous section can be computed numerically~\cite{Safa:2021ghs, Garg:2022ugd, Garcia:2020jwr}. However, in this work we use analytic estimates by approximating the PREM parametrization of the Earth density~\cite{Dziewonski:1981xy} by a set of thin, homogeneous layers. For a given trajectory inside a spherical Earth with exit angle $\theta$, the particle changes layers at the positions
\begin{equation}
	x_i^\pm = R_\oplus\cos\theta \pm \frac{\sqrt{2}}{2}\sqrt{2r_i^2 -R_\oplus^2(1-\cos 2\theta)}\, , 
\end{equation}with $r_i$ the radii of the discontinuities between layers and $R_\oplus$ the radius of the Earth. The width of each layer is $\Delta l_i = x_{i+1}-x_i$. 

The probability for a particle $\Nu$ to interact with a nucleus in a medium after travelling a distance $x$ is $p_\lambda(x; \lambda) = e^{-x/\lambda}/\lambda$, where $\lambda^{-1} = n\sigma$ is the mean free path and $n$ the nucleon number density. The probability for $\Nu$ to leave a uniform medium of depth $\Delta l$ is $
	P(X_{\text{int}}> \Delta l) = \int_{\Delta l}^\infty  p_\lambda(x; \lambda)\,\mathrm{d}x = e^{-\Delta l/\lambda}\,$, and the probability to escape all the layers is
\begin{equation}
	P_{\text{exit}}^\Nu = \prod_{i=1}^{m} P(X_{\text{int}}^{(i)} > \Delta l_i) = \prod_{i=1}^{m}  e^{-\Delta l_i/\lambda_i}\, ,
\end{equation}with $m$ the number of layers crossed in the trajectory.
In order to account for $\Nu$ regeneration, we add two additional terms, 
\begin{equation}
\begin{split}
	P_{\text{exit}}^\Nu =& 
	\prod_{i=1}^{m} P(X_{\text{int}}^{(i)} > \Delta l_i) 
	+ \\ &+ 
	\sum_{i=1}^m\underbrace{ \left(\prod_{k<i}P(X_{\text{int}}^{(k)}>\Delta l_k)\right)}_{\Nu \text{ survives all layers before } i} 
	\times
	\underbrace{ \left(\prod_{k>i}P(X_{\text{int}}^{(k)}>\Delta l_k)\right)}_{\Nu\text{ survives all layers after }i}
	\times\\&\ \times 
	\underbrace{\int_0^{\Delta l_i}P(X_{\text{int}}^{(i)} = x)\,\mathrm{d}x\int_x^{\Delta l_i}P(Y_{\text{dec}}^{(i)}>y-x) P(Y_{\text{int}}^{(i)}=y-x) 
		P(X_{\text{int}}^{(i)}>\Delta l_i-y)\, \mathrm{d}y}_{\textrm{In the layer }i,\ \Nu\text{ produces a }\Tau\text{ that produces another }\Nu} 
	+ \\ &+ 
	\sum_{i=1}^m\sum_{j>i}^m \underbrace{\left(\prod_{k<i}P(X_{\text{int}}^{(k)}>\Delta l_k)\right) }_{\Nu\text{ survives all layers before }i} 
	\times
	\underbrace{ \left(\prod_{k>j}P(X_{\text{int}}^{(k)}>\Delta l_k)\right)}_{\Nu\text{ survives all layers after }j} 
	\ \times \\&\ \times
	\underbrace{\int_0^{\Delta l_i} P(X_{\text{int}} = x)P(Y_{\text{decay}} > \Delta l_i-x)P(Y_{\text{int}} > \Delta l_i-x)\,\mathrm{d}x}_{\Nu \text{ interacts at layer }i\text{ and }\Tau\text{ leaves the layer}} 
	\ \times \\&\ \times
	\underbrace{ \left(\prod_{k=i+1}^{j-1}P(Y_{\text{decay}}^{(k)}>\Delta l_k)P(Y_{\text{int}}^{(k)}>\Delta l_k)\right)}_{\Tau\text{ survives all layers between $i$ and $j$}}
	\ \times \\&\ \times
	\underbrace{\int_0^{\Delta l_j} P(Y_{\text{decay}}^{(j)}>y)P(Y_{\text{int}}^{(j)}=y)
		P(X_{\text{int}}^{(j)}>\Delta l_j-y)\, \mathrm{d}y}_{\Tau\text{ produces a }\Nu\text{ in layer $j$ and $\Nu$ leaves}}
\end{split}  
\end{equation}
The first term describes $\Nu$ exiting without interaction. The second term introduces one intermediate $\Tau$ which is created and destroyed in the same layer. In the third term, the $\Tau$ is created and destroyed in different layers. This gives the following more concise result

\begin{equation}
\begin{split}
    P^\Nu_{\text{exit}} = \left(\prod_{k=1}^{m} e^{-\Delta l_k/\lambda_k}\right)\biggr[\, 
    1 & \ +
    \sum_{i=1}^m \left(\frac{c\tau}{\lambda_i}\right)^2 \left(\frac{\Delta l_i}{c\tau} + e^{-\Delta l_i/c\tau}-1\right)
	\ + \\&\ +
    \sum_{i,j>i}^m \frac{(c\tau)^2}{\lambda_i\lambda_j}\left(1-e^{-\Delta l_i/c\tau}\right)\left(1-e^{-\Delta l_j/c\tau}\right)
    \biggr]
\end{split}
\end{equation}We find that one regeneration process is enough to describe the dominant contributions.

\subsection{T exit probability}
In order for a $\Tau$ particle to exit a medium, we need the parent $\Nu$ to interact, and $\Tau$ not to decay nor interact before it leaves the medium. 
We define
\begin{equation}
	P(Y_{\text{decay}}>\Delta x) \equiv \int_{\Delta x}^\infty p_\tau(x,\tau)\,\mathrm{d}x = e^{-\Delta x/c\tau}\, ,\ 
	P(Y_{\text{int}}>\Delta x) \equiv \int_{\Delta x}^\infty p_\lambda(x,\tau)\,\mathrm{d}x = e^{-\Delta x/\lambda}\, .
\end{equation}
Treating Earth as a multi-layered medium, the total exit probability is
\begin{equation}
\begin{split}
	P_{\text{exit}}^\Tau = \sum_{i=1}^m&\underbrace{ \left(\prod_{j<i}P(X_{\text{int}}^{(j)}>\Delta l_j)\right)}_{\Nu\text{ survives layers before }i}\times \underbrace{\left(\prod_{j>i}P(Y_{\text{decay}}^{(j)}>\Delta l_j)P(Y_{\text{int}}^{(j)}>\Delta l_j)\right)}_{\Tau \text{ leaves all layers after }i} \\ &\times\underbrace{\int_0^{\Delta l_i} P(X_{\text{int}} = x)P(Y_{\text{decay}} > \Delta l_i-x)P(Y_{\text{int}} > \Delta l_i-x)\,\mathrm{d}x}_{\Nu\text{ interacts at layer }i\text{ and $\Tau$ leaves the layer}}\, .
\end{split}
\end{equation}This gives
\begin{equation}
	P_{\text{exit}}^\Tau = \left(\prod_{j=1}^{m}e^{-\Delta l_j/\lambda_j}\right)\left(\prod_{j>i}e^{-\Delta l_j/c\tau}\right)\sum_{i=1}^m \frac{c\tau}{\lambda_i}\left(1-e^{-\Delta l_i/c\tau}\right)\, .
\end{equation}

In ANITA, the total travelled distance is the chord length inside Earth, $L(\theta) = 2R_\oplus\cos\theta$. In IceCube the trajectory must finish at the detector, before leaving Earth. The distance between the IceCube detector and the exit point is 
\begin{equation}\label{eq:IC-dist-to-exit}
	2R_\oplus^2[1-\cos(\theta-\alpha(\theta))] 
	- 2R_\oplus D[1-\cos(\theta-\alpha(\theta))] + D^2 \equiv a(\theta) \, ,
\end{equation}where $D = 1\textrm{ km}$ is the depth of the IceCube detector. The trajectory is thus finished when the total travelled distance is $L(\theta)-a(\theta)$.

\section{Details of the test statistic}\label{app:ts}
In this section, we describe our statistical analysis. As data is scarce, an unbinned Poisson likelihood is well-suited. This test statistic is given, up to constants, by
\begin{equation}
    \mathcal{TS}(\Phi,\sigma,\tau) =
2\int \mathrm{d}\varphi\, \mathrm{d}\theta\sin\theta\, \mu(\theta,\varphi;\Phi,\sigma,\tau)
- 2\sum_{i=1}^N\log\tilde{\mu}(\theta^{\text{rec}}_i;\Phi,\sigma,\tau)  
\, ,
\end{equation}
where ${\mu \equiv dN/d\Omega}$, $\tilde{\mu}(\theta^{\text{rec}}) \equiv \sin\theta^{\text{rec}} \int_0^{2\pi} d\varphi\,  d\bar{N}/d\Omega(\theta^{\text{rec}},\varphi)$, and $\theta_i^\mathrm{rec}$ are the reconstructed angles of the $N$ observed events.

Since the ANITA-IV flight has detected 4 anomalous events, 
\begin{equation}\label{eq:test_statistic-anita}
\mathcal{TS}^{\text{ANITA}}(\Phi,\sigma,\tau) =
2\int \mathrm{d}\varphi\, \mathrm{d}\theta\sin\theta\, \mu(\theta,\varphi;\Phi,\sigma,\tau)
- 2\sum_{i=1}^4\log\tilde{\mu}(\theta^{\text{rec}}_i;\Phi,\sigma,\tau)  
\, .
\end{equation} 
In turn, IceCube has not observed any event in the energy range of the ANITA anomalous events. Then,
\begin{equation}\label{eq:test_statistic_ic}
	\mathcal{TS}^{\text{IC}}(\Phi,\sigma,\tau) = 2\int \mathrm{d}\varphi\, \mathrm{d}\theta\sin\theta\, \mu(\theta,\varphi;\Phi,\sigma,\tau) \, .
\end{equation}This test statistic is twice the total number of expected events in IceCube. The bigger the expected number of events, the worse the fit.

For an isotropic flux, $f_\Omega(\Omega) = (4\pi)^{-1}$. Then, integrating over $\varphi$
\begin{equation}\label{eq:ts-ANITA-diffuse-withphi} 
\begin{split}
 \mathcal{TS}^{\textrm{ANITA}}(\Phi,\sigma,\tau) = & 
 -2\sum_{i=1}^4\log\left(\frac{\Phi\,\Delta t}{2}\sin\theta^{\text{rec}}_i\, \overline{\mathcal{A}}(\theta^{\text{rec}}_i,\Delta \theta_i;\sigma,\tau)\right) + \\ & + \Phi\,\Delta t\int \mathrm{d}\theta\, \sin\theta\, \mathcal{A}(\theta;\sigma,\tau)\, .
\end{split}
\end{equation}
Here we have defined
\begin{equation}
	\overline{\mathcal{A}}(\theta^{\text{rec}},\Delta \theta;\sigma,\tau) = 
	\int \mathrm{d}\theta \, \mathcal{A}(\theta;\sigma,\tau)  
	\frac{1}{\sqrt{2\pi}\Delta\theta} \exp\left[ -\frac{(\theta^{\text{rec}}-\theta)^2}{2(\Delta\theta)^2} \right]\, .
\end{equation}
The best-fit value of $\Phi$ for this likelihood is
\begin{equation}\label{eq:ts-ANITA-diffuse-phbestfit}
\Phi^{\textrm{BF}}_\mathrm{ANITA} = \frac{2 \times 4}{\Delta t \int \mathrm{d}\theta\, \sin\theta\,  \mathcal{A}(\theta;\sigma,\tau)}\, ,
\end{equation}
that only depends on the total number of observed and expected events. 
We can then insert \Cref{eq:ts-ANITA-diffuse-phbestfit} in \Cref{eq:ts-ANITA-diffuse-withphi} to obtain the ANITA test statistic profiled over $\Phi_\mathrm{ANITA}$,
\begin{equation}\label{eq:ts-ANITA-diffuse-nophi}
\mathcal{TS}^{\textrm{ANITA}}(\sigma,\tau) = 
\sum_i - 2\log\left(\frac{\sin\theta^{\text{rec}}_i\,\overline{\mathcal{A}}(\theta^{\text{rec}}_i;\sigma,\tau)}{\int \mathrm{d}\theta\, \sin\theta\, \mathcal{A}(\theta;\sigma,\tau)}\right)\, ,
\end{equation}up to constant terms. The best-fit values of $(\sigma,\tau)$ must maximize the argument of the logarithm. That is, they must concentrate the flux around $\theta_i^{\text{rec}}$. This way, the probability for the events to happen at $\theta_i^{\text{rec}}$ and not elsewhere is maximal. 

Again, the IceCube test statistic is quite simple,
\begin{equation}\label{eq:ts-IC-diffuse}
	\mathcal{TS}^{\text{IC}}(\Phi,\sigma,\tau) = \Phi\,\Delta t\int \mathrm{d}\theta \sin\theta \mathcal{A}(\theta;\sigma,\tau) \, .
\end{equation}

The total test statistic for both experiments contains the likelihoods from~\cref{eq:ts-ANITA-diffuse-withphi,eq:ts-IC-diffuse},
\begin{equation}\label{eq:ts-total-diffuse} 
\begin{split}
 \mathcal{TS}(\Phi,\sigma,\tau) = & 
 -2\sum_{i=1}^N\log\left(\frac{\Phi\,\Delta t_{\mathrm{A}}}{2}\,\sin\theta^{\text{rec}}_i\, \overline{\mathcal{A}}_{\mathrm{A}}(\theta^{\text{rec}}_i,\Delta \theta_i;\sigma,\tau)\right) + \\ & + 
 \Phi\,\Delta t_{\mathrm{A}}\int \mathrm{d}\theta\, \sin\theta\, \mathcal{A}_{\mathrm{A}}(\theta;\sigma,\tau)+\Phi\,\Delta t_{\mathrm{I}}\int \mathrm{d}\theta\, \sin\theta\, \mathcal{A}_{\mathrm{I}}(\theta;\sigma,\tau) \, ,
\end{split}
\end{equation}where the A, I subscripts stand for ANITA and IceCube, respectively. This test statistic can be analytically profiled over $\Phi$, which gives
\begin{equation}\label{eq:ts-GF-nophi}
	\mathcal{TS}(\sigma,\tau) = 
	- 2\sum_{i=1}^{4}\log\left(
        \frac{
              \sin\theta^{\text{rec}}_i\, \overline{\mathcal{A}}_\mathrm{A}(\theta^{\text{rec}}_i;\sigma,\tau)
              }{
              \Delta t_\mathrm{A}\, \mathcal{A}^{\text{tot}}_{\text{A}}+ 
              \Delta t_\mathrm{I}\, \mathcal{A}^{\text{tot}}_{\text{I}}
              }
        \right)\, , 
\end{equation}up to constant terms. We have abbreviated $\mathcal{A}^{\text{tot}} = \int \mathrm{d}\theta\, \sin\theta\, \mathcal{A}(\theta;\sigma,\tau)$. The best-fit flux is
\begin{equation}
	\Phi(\sigma,\tau)^\mathrm{BF} = \frac{2 \times 4}{ \Delta t_\mathrm{A}\, \mathcal{A}^{\text{tot}}_{\text{A}}+ 
              \Delta t_\mathrm{I}\, \mathcal{A}^{\text{tot}}_{\text{I}} }\, .
\end{equation}

\end{document}